\documentclass[preprint,12pt]{elsarticle}

\usepackage{amssymb}
\usepackage{dcolumn,amsmath}
\usepackage{bm}
\usepackage{multirow}

\usepackage{amsfonts}
\usepackage{cancel}

\usepackage{color}

\journal{Chemical Physics Letters}

\begin{document}

\begin{frontmatter}



\title{Rovibrational structure and electric dipole moments of the AcOCH$_3$+ ion}


\author[label1,label2]{Anna Zakharova}

\ead{zakharova.annet@gmail.com}

\affiliation[label1]{organization={St. Petersburg State University},
            addressline={7/9 Universitetskaya nab.}, 
            city={St. Petersburg},
            postcode={199034}, 
            state={},
            country={Russia}}

\affiliation[label2]{organization={Petersburg Nuclear Physics Institute named by B.P. Konstantinov of National Research Centre
"Kurchatov Institute"},
            addressline={1, mkr. Orlova roshcha}, 
            city={Gatchina},
            postcode={188300}, 
            state={},
            country={Russia}}

\begin{abstract}
The possibility of  laser cooling and the presence of closely spaced rovibrational doublets make polyatomic molecules an attractive platform for the $\mathcal{P}$, $\mathcal{T}$-violation searches. We study the spectrum of the lowest rovibrational state of the AcOCH$_3+$ symmetric top molecule.
The electronic structure full-electron computation was performed within a relativistic coupled cluster method with double and perturbative triple excitations. The rovibrational wavefunctions are obtained using a coupled channel technique, taking into account all rovibrational effects and anharmonicities of the potential. As a result, the vibrational frequencies, as well as the values of the electric dipole moments for the rovibrational states, were computed.

\end{abstract}



\end{frontmatter}
\section{Inroduction}

Due to the lack of evidence for the New Physics at the Large Hadronic Collider, other experimental methods become important for the development of fundamental physics. The physics at very large energy scales can still produce an echo at low energies in the form of  fundamental parity violations such as space reflection ($\mathcal{P}$), charge conjugation ($\mathcal{C}$), and time reversal ($\mathcal{T}$) that may be stronger than the corresponding effects in the Standard Model \cite{schwartz2014quantum,particle2020review}. Thus, the discrete symmetry violations may serve as a powerful window into the New Physics \cite{khriplovich2012cp}. 
As the only source of such violation in the Standard Model is the complex-valued Cabibbo-Kobayashi-Maskawa (CKM) \cite{Cabibbo1963,KobayashiMaskawa1973} and Pontecorvo–Maki–Nakagawa–Sakata (PMNS)  \cite{Pontecorvo1957,MNS1962} mixing matrices, some effects are strongly suppressed compared to various scenarios of the New Physics. Examples include the electron electric dipole moment (eEDM) and scalar-pseudoscalar electron-nucleon interaction (S-PS) \cite{Fukuyama2012,PospelovRitz2014,YamaguchiYamanaka2020,YamaguchiYamanaka2021}.

The strongest limits on such effects are obtained in molecular experiments \cite{ginges2004violations,PospelovRitz2014,ChubukovLabzowsky2016, baron2014order,ACME:18}.
 The sensitivities to such $\mathcal{P}$, $\mathcal{T}$-violating effect have to be obtained in quantum chemical calculations \cite{KozlovLabzowsky1995, titov2006d, Safronova2017}.
The best current EDM limits were received in diatomic molecules HfF$^{+}$ \cite{Cornell:2017,Petrov:18,roussy2023improved} and ThO\cite{ACME:18,DeMille:2001,Petrov:14,Vutha:2010,Petrov:15,Petrov:17}. The power of these experiment comes from the presence of the electronic levels of opposite parity in their spectrum, allowing to reduce systematic effects by measuring different transitions.

The polyatomic molecules also admit the rovibrational states such as $l$-doublets and, in case of the symmetric top molecules, $K$-doublets. However, compared to the diatomics, they have advantage of the possibility of the laser cooling \cite{kozyryev2017sisyphus,steimle2019field,augenbraun2020laser}. However, due to the rovibrational nature of these states, the influence of vibrations and rotations on sensitivity to eEDM and S-PS interactions must be taken into account \cite{gaul2020ab,ourRaOH,zakharova2021rovibrational}.

One of the most promising types of the molecules for this task are symmetric top molecules with general formula (M)OCH$_3$, such as RaOCH$_3$ \cite{zakharova2022rotating}, and YbOCH$_3$ that also allow laser-cooling \cite{isaev2016polyatomic, kozyryev2016proposal,kozyryev2019determination,augenbraun2021observation}.
In addition to the states with nonzero rovibrational momentum $l$ such as in the triatomic molecules they also have the doublets associated with the nonzero projection of the total angular momentum $K$ on the molecular axis. Laser cooling of another symmetric top, CaOCH$_3$, was demonstrated perspective in  \cite{mitra2020direct}. In \cite{fan2021optical} the  controlled syntheses of cold symmetric-top ion RaOCH$_3^+$ was achieved. 

Without the external field the states of the molecules are not sensitive to the parity violation effects, as they are parity eigenvalues such as,
\begin{equation}
|\pm\rangle = \frac{1}{\sqrt{2}}\Big(|+l\rangle \pm |-l\rangle\Big),
\end{equation}
and similar states with $\pm K$ in case of the $K$ doublets. However, in the external electric field these states polarize and gain a shift,
\begin{equation}
    \delta E = P (d_eE_{eff}+k_sE_s)
\end{equation}
The strength of the electric field required for the full polarization is determined from the gap between the $|\pm\rangle$ energy level as well as on the value of the electric dipole moments of the molecule in these states. As well, the determination of the polarization coefficient $P$ requires taking into account a number of the effects associated with the sensitivities of the spectra to various perturbation. In this work we determine the rovibrational spectrum of the lowest states for the AcOCH$_3+$ symmetric top molecular ion, as well as compute the electric dipole momenta averaged over the rovibrational wavefunction.

\section{The coupled channel method}
\label{Sec:Geometry}

\begin{figure}[h]
\centering
\label{Sec:СС}
\includegraphics[width=0.5\textwidth]{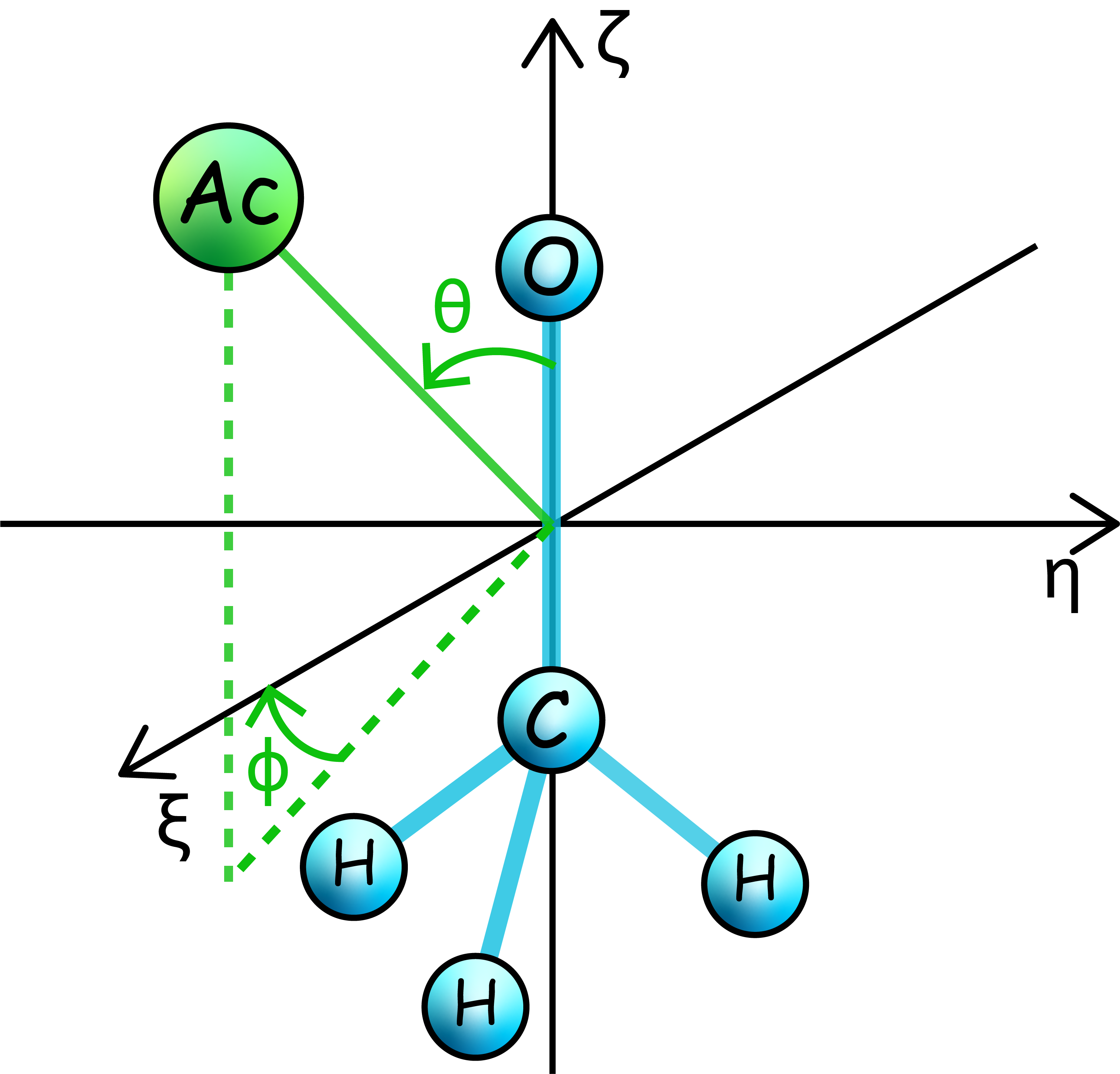}
  \caption{The AcOCH$_3$+ molecule}
  \label{fig:molecule}
\end{figure}

In the Born-Oppenheimer approximation the wavefunction can be represented as a factor of the electronic and nuclear motion parts,
\begin{equation}
\Psi_{\rm total}\simeq \Psi_{\rm nuc}(R,\hat{R},\hat{r},\gamma)\psi_{\rm elec}(\{\vec{r}_i\}|R,\theta,\varphi),
\label{psiexp}
\end{equation}
where $R$, $\theta$ and $\varphi$ define the geometry of the AcOCH$_3$+ molecule in Fig.~\ref{fig:molecule}. The angle $\gamma$ determines the orientation of the CH$_3$ around $\zeta$. The unit vectors along Yb -- ligand c.m. axis, $\hat{R}$, and ligand $\zeta$ axis, $\hat{r}$, also determine the configuration. 

Treating the OCH$_3$ ligand as rigid, we consider the AcOCH$_3$+ ion as a coupled two-body system. The first part is a linear rotor of length $R$, comprising the Yb atom and the ligand's center of mass. Its rotational motion is described by the angular momentum operator $\hat{\vec{l}}$, with quantum numbers $l$ (for $\hat{\vec{l}}^2$) and 
and $m_l$ (its projection onto the z-axis of the space-fixed frame, SFF). The second component is the OCH$_3$ ligand itself, treated as a rigid symmetric top with angular momentum $\hat{\vec{j}}$. This rotor is characterized by the quantum numbers: $j$ for $\hat{\vec{j}}^2$, $m$ for its projection on the $z$ axis, and $k$ for its projection on the $\zeta$ axis of the ligand-fixed frame.

\begin{equation}
    \hat{H}_{nuc }=\left[-\frac{1}{2 \mu} \frac{\partial^2}{\partial R^2}+\frac{\hat{\vec{l}}^2}{2 \mu R^2}+V \left( R, \hat{R}, \hat{r}, \gamma\right)\right]+\hat{H}_{lig},
\end{equation}

\begin{equation}
\hat{H}_{\text{lig}}=\frac{\hat{j}_{\xi}}{2 I_{\xi}}+\frac{\hat{j}_{\eta}}{2 I_{\eta}}+\frac{\hat{j}_{\zeta}^2}{2 I_{\zeta}},
\end{equation}
Here, $\xi,\eta, \zeta $ denote the principal axes of the ligand-fixed coordinate system.

\begin{equation}
\hat{H}_{\text{nuc}} \psi^{JM}(R,\hat{R},\hat{r},\gamma)=E \psi^{JM}(R,\hat{R},\hat{r},\gamma),
\end{equation}
where the unit vectors $\hat{R}$ and $\hat{r}$, together with the angle $\gamma$, specify the orientation of the molecule in the space-fixed frame.

The continuous angular dependence of $\psi^{JM}$ can be discretized by expanding it in a basis of angular momentum eigenfunctions $\hat{\vec{j}}^2, \hat{j}_\zeta$ and $\hat{\vec{l}}^2$:
\begin{equation}
\psi^{J M}(R, \hat{R}, \hat{r}, \gamma)=\sum_{j=0}^{\infty} \sum_{k=-j}^{+j} \sum_{l=0}^{+\infty} F_{j k l}^J(R){Y}_{j k l}^{JM}(\hat{R}, \hat{r}, \gamma).
\end{equation}

\begin{equation}
    \Big(\frac{d^2}{d R^2} -\frac{l(l+1)}{ R^2} +2\mu E - 2\mu E_{\rm lig}^{jk}\Big)F_{j k l}^J(R) \\
    \\= 2 \mu \sum_{j'k'l'} v_{jklj'k'l'}(R)F_{j' k' l'}^J(R),
\end{equation}

\begin{multline}
v_{jkl,\tilde{j}\tilde{k}\tilde{l}}(R)=(-1)^{j+\tilde{j}+\tilde{k}-J}\sqrt{\frac{(2j+1)(2\tilde{j}+1)(2l+1)(2\tilde{l}+1)(2\lambda+1)}{4\pi}}\\
\times\sum_{\lambda\mu} V_{\lambda\mu}(R)
\begin{Bmatrix}j&l&J\\\tilde{l}&\tilde{j}&\lambda\end{Bmatrix}\begin{pmatrix}l&\lambda&\tilde{l}\\0&0&0\end{pmatrix}\begin{pmatrix}j&\lambda&\tilde{j}\\-k&\mu&\tilde{k}\end{pmatrix},
\end{multline}
with $V_{\lambda\mu}(R)$ obtained by expanding the adiabatic potential,
\begin{equation}
V(R, \hat{R}, \hat{r})=V(R, \theta, \varphi)=\sum_{\lambda=0}^{+\infty} \sum_{\mu=-\lambda}^{+\lambda} V_{\lambda \mu}(R) Y_{\lambda \mu}( \theta, \varphi).\label{spherical_decompose}
\end{equation}

The only continuous variable that remain is the radial coordinate $R$. To discretize it we decompose all the functions on $R$ in terms of the eigenfunctions $f_n(R)$ for longitudinal oscillations in the harmonic approximation:
\begin{equation}
\Big[-\frac{1}{2\mu}\frac{d^2}{dR^2}+\frac{\mu\omega_\parallel^2}{2} (R-R_{eq})^2\Big] f_n(R) = \omega_\parallel\Big(n + \frac{1}{2}\Big)f_n(R).
\end{equation}

The operators are then represented as discrete matrices such as for the centrifugal term:
\begin{equation}
C_{n\tilde{n}}=\int_{R_{min}}^{R_{max}} dR\,f_n(R)\frac{1}{R^2}f_{\tilde{n}}(R),
\end{equation}
and for the potential:
\begin{equation}
\tilde{V}_{kn\tilde{n}}=\int_{R_{min}}^{R_{max}} dR\,f_n(R) \Big(V_k(R)-\delta_{k,0}\frac{\mu\omega_{\parallel}^2}{2}(R-R_{eq})^2\Big)f_{\tilde{n}}(R).\label{Integ}
\end{equation}
The values for $R_{min}$ and higher $R_{max}$ should be chosen in such a way that the obtained nuclear motion wavefunctions decay when they approach the boundaries of that range.

In practice the adiabatic potential surface is obtained on a discrete grid of the $(R, \theta, \phi)$. First we approximate the dependence on $R$ for $\theta=0^\circ$ using the Morse potential which allows us to obtain the parameters for the $f_n(R)$ eigenbasis. Then for each value of $R_k$ the decomposition \eqref{spherical_decompose} is obtained. The $R$-depenence of the coeffiecients is interpolated with Akima splines. Subsequently the numerical integrals of $v_{jklj'k'l'}(R)$ with $f_n(R)$ are computed to obtain the discrete matrix elements. The numerical stability is checked by varying the maximal limits for all quantum numbers in the decomposition.

\section{Application to AcOCH$_3+$}

The electronic computations were performed using DIRAC 19 \cite{DIRAC19} program suite. The self-consistent field computations were done using a four-component Kramers-restricted Dirac-Harthree-Fock method. For the Actinium atom a full-electron Dyall ae2z basis was used. The cc-pVTZ basis set was used for the Oxygen, Carbon and Hydrogen atoms. The electronic correlations were taken into account using the relativistic coupled cluster method with double and perturbative triple excitations (CCSD(T)).

The following grid of the Jacobi coordinates was used:
\begin{itemize}
    \item Evenly spaced grid of $R = 5.1, 5.3,\ldots ,5.9\,\mathrm{a.u.}$, $\theta=0, 5^\circ$;
    \item For $\theta <= 30^\circ$ we used small step $5^\circ$, followed by $15^\circ$ step for $\theta <= 90^\circ$ and extra $\theta=180^\circ$ point
    \item For $\phi$ we used evenly spaced grid with $15^\circ$ step, but only for $0^\circ \leq \phi \leq 60^\circ$ as other regions could be restored by the symmetry of the molecule.
\end{itemize}
This choice of grid allows to accurately describe the most important part of the potential for the states under consideration while keeping the overall amount of computations sufficiently low.

\begin{figure}[h]
\centering
  \includegraphics[width=0.5\textwidth]{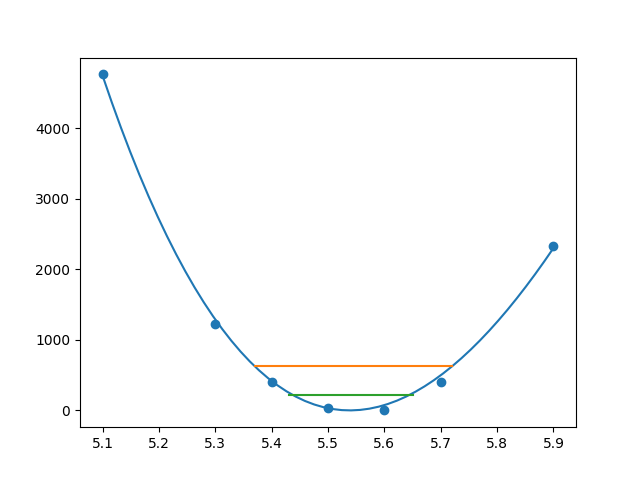}
  \caption{The dependence of the adiabatic potential (in $\mathrm{cm}^{-1}$) for $\theta=0^\circ$ on $R$ (in a.u.) The ground state and first longitudinally excited state energy levels are depicted as horizontal lines.}
  \label{fig:radial_potential}
\end{figure}
The dependence of the adiabatic potential on $R$ for the linear case is depicted on Fig.~\ref{fig:radial_potential}. As one can see the Morse potential with $R_e=5.54\,\mathrm{a.u.}$, $\omega=423.4\,\mathrm{cm}^{-1}$ and $D_e=116437\,\mathrm{cm}^{-1}$ provides a very good fit. Also, one can see that the wavefunction of the states considered in this work is concentrated in the region covered by our grid.

\begin{figure}[h]
\centering
\includegraphics[width=0.5\textwidth]{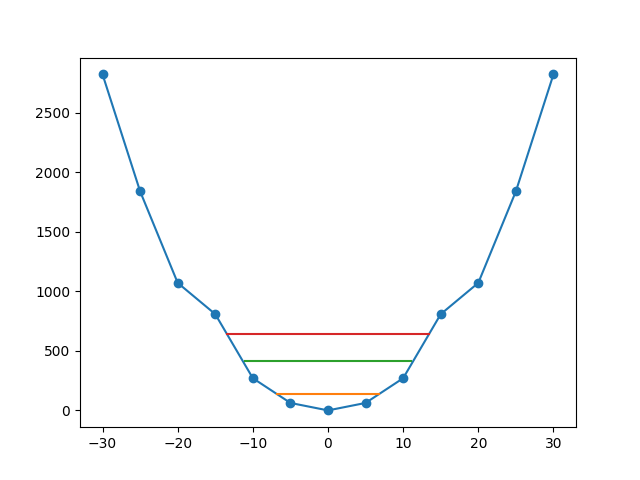}
  \caption{The dependence of the $\phi$-averaged adiabatic potential (in $\mathrm{cm}^{-1}$) for $R=R_{eq}$ on $\theta$ (in Degrees). The ground state and the lowest transversely excited state energy levels are depicted as horizontal lines.}
  \label{fig:angular_potential}
\end{figure}
The dependence of the $\phi$-averaged adiabatic potential on the $\theta$ for $R=R_{eq}$ is depicted on Fig.~\ref{fig:angular_potential}. Despite the states under consideration being concentrated near the equilibrium point, the impact of the anharmonicity, and the mixing with the rotational degrees of freedom becomes noticeable already for the second excited state.

\begin{figure}[h]
\centering
\includegraphics[width=0.5\textwidth]{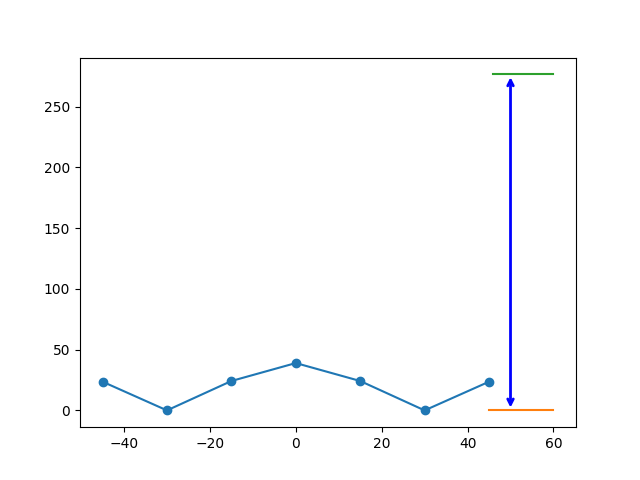}
  \caption{The dependence of the adiabatic potential (in $\mathrm{cm}^{-1}$) for $\theta=10^\circ$ on $\phi$ (in $\circ$) For comparison the distance between the ground state and first transversely excited energy level is depicted.}
  \label{fig:azimuthal_potential}
\end{figure}
The azimuthal dependence of the potential on $\phi$ for $R=R_{eq}$ and $\theta=10^\circ$ (the approximate extent of the wavefunctions for the lowest excited states) is depicted on Fig.~\ref{fig:azimuthal_potential}. As the variation of the potential with $\phi$ is rather small compared to the distance between the lowest energy levels, for most of the properties this can be considered as a perturbation. However, this dependence should be taken into account for such precise properties as $K$-doubling and $l$-doubling.

The longitudinal electric dipole moment of the molecule $d_\parallel$ behaves practically linear with $R$. The transverse electric dipole moment is averaged to zero for all considered states, though it should play important role in the mixing caused by the application of the transverse electric field.

\begin{table}[h]
\small
\centering
  \caption{The longitudinal electric dipole moment for various states of the AcOCH$_3+$}
  \label{tbl:results}
  \renewcommand{\arraystretch}{1.5}
  \begin{tabular*}{0.52\textwidth}{@{\extracolsep{\fill}}lllll}
    \hline\hline
$v_{\parallel}$ & $v_\perp$ &  $l$ &  $E-E_0,\,\mathrm{cm}^{-1}$ & $d_\parallel, \mathrm{a.u.}$\\
\hline
  \multicolumn{1}{l}{J=0}\\
\hline
0 & 0 & 0 & 0 & 7.61\\
1 & 0 & 0  & 402.30  & 7.50\\
0 & 2 & 0 & 503.46 & 7.37\\
2 & 0 & 0  & 813.99 & 7.37\\
\hline
  \multicolumn{1}{l}{J=1}\\
  \hline
0 & 1 & $\pm 1$ & 276.46  & 7.48\\
 \hline\hline
  \end{tabular*}
\end{table}

The energies of the various states, as well as the corresponding longitudinal electric dipole moment values, are given in Table~\ref{tbl:results}.


\section{Acknowledgement}
The work was supported by the Theoretical Physics and Mathematics Advancement Foundation “BASIS” (grant N\textsuperscript{\underline{o}}23-1-4-10-1  ).


\end{document}